\documentclass[aps,twocolumn,floatfix,a4]{revtex4}

\usepackage{amsmath}
\usepackage{latexsym}
\usepackage{float}
\usepackage{amssymb}
\usepackage{graphicx}
\usepackage{epsfig}
\usepackage{color}

\begin{document}
\title{Detailed analysis of Rouse mode and dynamic scattering function of
highly entangled polymer melts in equilibrium}
\author{Hsiao-Ping Hsu}
\email{hsu@mpip-mainz.mpg.de}
\author{Kurt Kremer}
\email{kremer@mpip-mainz.mpg.de}
\affiliation{Max-Planck-Institut f\"ur Polymerforschung, 55128 Mainz, Germany}
\begin{abstract}
We present large-scale molecular dynamics simulations for a coarse-grained
model of polymer melts in equilibrium. 
From detailed Rouse mode analysis we show that 
the time-dependent relaxation of the autocorrelation function (ACF) of modes $p$
can be well described by the effective stretched exponential function due to the 
crossover from Rouse to reptation regime.
The ACF is independent of chain sizes $N$ for $N/p<N_e$ ($N_e$ is the entanglement length),
and there exists a minimum of the stretching exponent as $N/p \rightarrow N_e$.
As $N/p$ increases, we verify the crossover scaling behavior of the effective 
relaxation time $\tau_{{\rm eff},p}$ from the Rouse regime to the 
reptation regime. We have also provided evidence that the incoherent dynamic
scattering function follows the same crossover scaling behavior of the mean square
displacement of monomers at the corresponding characteristic time scales.
The decay of the coherent dynamic scattering function is slowed
down and a plateau develops as chain sizes increase
at the intermediate time and wave length scales. The tube diameter extracted 
from the coherent dynamic scattering function is equivalent to the previous
estimate from the mean square displacement of monomers.

\end{abstract}
%
\maketitle
\section{Introduction}
\label{intro}

 The dynamics of polymer chains in a melt is a complicated many-body problem where
the motion of chains depends not only on different length scales but also time scales.
It is well known that for short unentangled chains in a melt, excluded volume
and hydrodynamic interactions are screened, and the viscoelastic properties of
chains can be approximately described by the Rouse model~\cite{Rouse1953,deGennes1979,Doi1986}.
If the polymer chains become long enough, the topological constraints
dominate the dynamics of the chains. At intermediate time and length scales,
the chains are assumed to move back and forth (reptation) within a tube-like region 
created by surrounding entangled chains and depending on the
entanglement length $N_e$. The dynamic behavior within this
time frame is well described by the reptation theory of de Gennes,
Doi and Edwards~\cite{deGennes1979,Doi1986,Rubinstein2003}.

  Rouse mode analysis provides a straightforward way of understanding the
dynamics of single chains in a melt by mapping the trajectories of the chains
into orthogonal Rouse modes. The chains are assumed to 
be Gaussian in this analysis. This has been applied widely to the analysis of 
experimental data and simulation data. Besides the original predictions of this model, 
in the literature~\cite{Shaffer1995,Padding2002,Kreer2001,Wittmer2007,Semenov2010}  
there also exist modified theoretical predictions 
including the excluded volume interaction,
the intrinsic stiffness of chains, the intramolecular correlations in
chains, and the topological constraints  
for analyzing the relaxation of Rouse modes.

In our recent work~\cite{Hsu2016}, we have investigated 
the chain conformations of fully equilibrated and highly entangled polymer 
melts and compared our simulation
results to the related theoretical predictions in detail. 
The fully equilibrated and highly entangled polymer
melts were generated by a novel and very efficient methodology 
through a sequential backmapping of soft-sphere coarse-grained configurations from
low resolution to high resolution, and finally the application of molecular dynamics (MD)
simulations of the underlying bead-spring model~\cite{Zhang2014,Moreira2015,Vettorel2010,Kremer1990}.
We have also studied the dynamics of fully equilibrated polymer 
melts, characterized by the mean square displacement of monomers,
and determined the characteristic time scales:
the characteristic time $\tau_0$, the entanglement time $\tau_e \approx \tau_0N_e^2$, 
the Rouse time $\tau_R\approx \tau_0N^2$, and the disentanglement time $\tau_d \approx \tau_0N^2(N/N_e)^{1.4}$,
according to the predictions given by the Rouse model and the reptation theory, where
$N_e$ is the entanglement length and $N$ is the chain size.
For $N<N_e$ in the Rouse regime, there exist exact solutions of almost all physical observables.
Therefore, based on this work, we are interested in understanding to 
what extent the dynamics of single chains
can be analyzed through the Rouse mode analysis
and check the scaling predictions of the relaxation of the Rouse modes
in the 
literature~\cite{Rouse1953,deGennes1979,Doi1986,Rubinstein2003,Shaffer1995,Padding2002,Kreer2001,Wittmer2007,Semenov2010,Kremer1990,Puetz2000,Kalathi2014}
whenever it is possible. 
On the other hand, Rouse mode decay should display a similar
slowing down of the dynamic structure factor.
The tube diameter introduced in the reptation theory can
be extracted from the single chain dynamic structure factor measured via neutron
spin echo (NSE) experiments~\cite{Wischnewski2000,Wischnewski2002,Wischnewski2003}.
Therefore we also study the dynamic scattering function from single 
chains in a melt, and check the consistency of the tube diameter estimated 
from different physical quantities~\cite{Doi1986,Kremer1990,Puetz2000,deGennes1981,kremer1984}. 
We use the ESPResSo++ package~\cite{Espresso} to perform the standard MD
simulations with Langevin thermostat at the temperature $T=1\varepsilon/k_B$ where
$k_B$ is the Boltzmann factor to study fully equilibrated 
polymer melts consisting of $1000$ chains of 
sizes $N=500$, $2000$ for $k_\theta=1.5$ ($\tau_0 \approx 2.89\tau$, $N_e \approx 28$~\cite{Hsu2016,Moreira2015})
and of sizes $N=1000$, $2000$ for $k_\theta=0.0$ ($\tau_0 \approx 1.5\tau$, $N_e \approx 87$~\cite{Moreira2015}) 
in the framework of the standard 
bead-spring model~\cite{Kremer1990} with a bond bending interaction parameter $k_\theta$
at a volume fraction $\phi=0.85$. 

The outline of the paper is as follows: 
Sec. II describes the theoretical background of
the Rouse model and the Rouse mode analysis of highly entangled polymer melts.
Sec. III describes the scaling behavior of dynamic structure factors and the comparison
between theory and simulation. Finally, our conclusions are summarized in Sec. IV.

\section{Rouse mode analysis}
In the Rouse model, neglecting inertia effects Rouse chains undergo
Brownian motion and therefore 
the Langevin equation of motion for the $i$th monomer is thus given by
\begin{equation}
  \zeta \frac{d \vec{r}_i}{dt}=
\frac{\partial U(\vec{r}_0,\ldots,\vec{r}_{N-1})}{\partial \vec{r}_i}
+\vec{f}_i^R(t)
\end{equation}
with the Rouse potential 
\begin{equation}
U(\vec{r}_0,\ldots,\vec{r}_{N-1})=\frac{3k_BT}{2b^2}
\sum_{i=1}^{N-1} (\vec{r}_{i+1}-\vec{r}_i)^2
\end{equation}
where  $b$ is the
effective bond length, $\vec{r}_i$ is the position vector of the $i$th monomer, and
$\vec{f}^R_i(t)$ is a random force. 
The friction $\zeta$ and the random 
force $\vec{f}_i^R(t)$ are related by the fluctuation dissipation theorem, 
i.e.,
\begin{equation}
 \langle \vec{f}_i^R(t) \cdot \vec{f}_j^R(t') \rangle = 
6 k_BT \zeta \delta_{ij} \delta(t-t') \, .
\end{equation}
The Rouse modes $\vec{X}_p(t)$ are defined as the cosine transforms 
of position vectors $\vec{r}_i$ at time $t$ for $i=1,\ldots,N$ as given in 
Ref.~\cite{Kopf1997},
\begin{eqnarray}
  \vec{X}_p(t)=\left(\frac{2}{N}\right)^{1/2}
\sum_{i=1}^{N} \vec{r}_i(t) \cos \left[\frac{p\pi}{N}(i-1/2)\right] \,,
\nonumber \\
p=0,\ldots,N-1 \,.
\label{eq-xpdef}
\end{eqnarray} 
Here the $p>0$ modes describe the internal relaxation
of a chain of $N/p$ monomers while the $0$th ($p=0$) mode 
corresponds to the motion of the center-of-mass of chain.
Since each Rouse mode $\vec{X}_p(t)$ for $p>0$ performs
a Brownian motion in a harmonic potential, independent of
each other, the cross-correlations should vanish,
the normalized autocorrelation function is expected to
follow an exponential decay,
\begin{equation}
   \frac{\langle \vec{X}_p(t) \vec{X}_p(0) \rangle}
{\langle \vec{X}_p(0) \vec{X}_p(0) \rangle}=\exp(-t/\tau_p)
\label{eq-Rouse}
\end{equation}
with 
\begin{eqnarray}
    \langle X_p^2 \rangle =\langle \vec{X}_p(0) \vec{X}_p(0) \rangle &=&b^2
\left[4 \sin^2 \left(\frac{p\pi}{2N} \right) \right]^{-1} \nonumber \\
&&\mathrel{\mathop{\longrightarrow}^{p/N \ll 1}}  b^2 \left(\frac{p\pi}{N}\right)^{-2}\,,
\label{eq-xpxp0}
\end{eqnarray}
and 
\begin{equation}
   \tau_p= \tau_0 \left(\frac{p}{N} \right)^{-2}=
\frac{\zeta b^2}{3k_BT \pi^2}\left(\frac{p}{N} \right)^{-2} \,.
\label{eq-taup}
\end{equation}
For $p=N$, $\tau_N=\tau_0$ is the shortest relaxation time of the Rouse model,
where $\tau_0=\zeta b^2/(3k_BT \pi^2)$ is the characteristic relaxation time,
while $\tau_1$ for $p=1$ is the longest relaxation time equal to
the Rouse time, i.e. $\tau_1=\tau_R$.

However, as the excluded volume interaction and topological
constraint are taken into account, it has been
pointed out in the literature~\cite{Shaffer1995,Padding2002,Li2012}
that simulation
results of the normalized autocorrelation function 
are well described by the stretched exponential 
Kohlrausch-Williams-Watts (KWW) function, i.e.
\begin{equation}
   \frac{\langle \vec{X}_p(t) \vec{X}_p(0) \rangle}
{\langle \vec{X}_p(0) \vec{X}_p(0) \rangle}=\exp 
[-(t/\tau_p^*)^{\beta_p} ] \,,
\label{eq-kww}
\end{equation}
where the KWW characteristic relaxation time $\tau_p^*$ and 
the stretching exponent $\beta_p$ depend on the mode $p$, and both are
measures of importance of excluded 
volume interactions and topological constraints.
The effective Rouse time of mode $p$ is thus given by
\begin{equation}
  \tau_{{\rm eff},p}=\int_0^\infty dt \exp [-(t/\tau_p^*)^{\beta_p} ]
= \frac{\tau_p^*}{\beta_p} \Gamma \left( \frac{1}{\beta_p}\right) \,,
\label{eq-taueff}
\end{equation}
where $\Gamma(x)$ is the gamma function.

\begin{figure*}[t]
\resizebox{1.60\columnwidth}{!}{\includegraphics{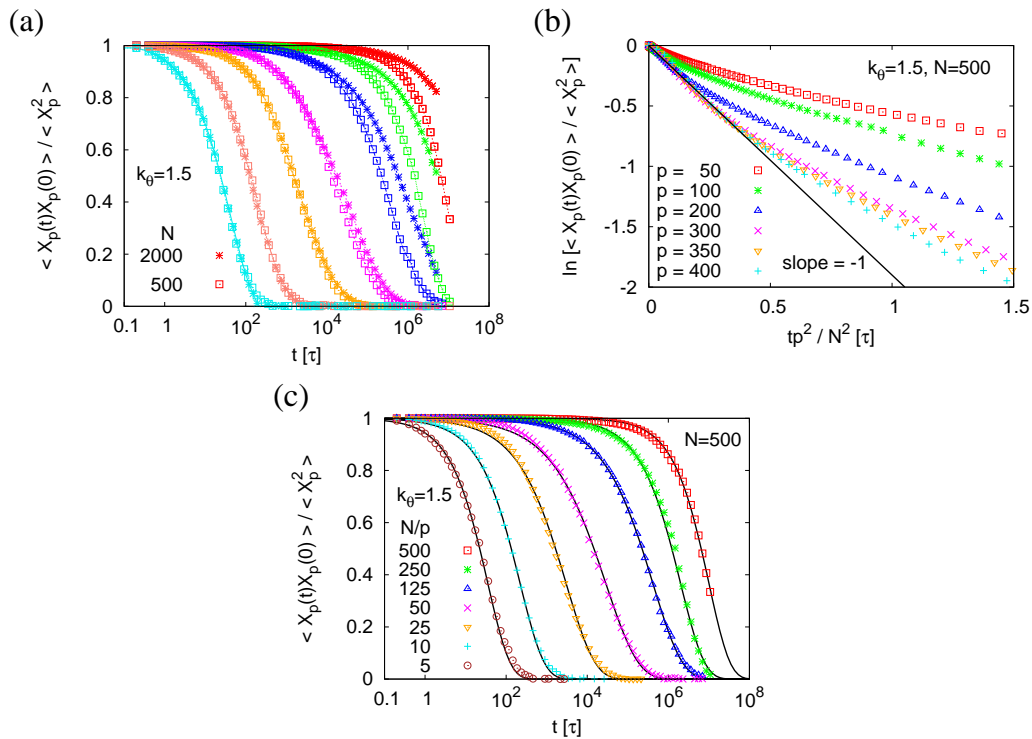}}
\caption{(a) Semi-log plot of the normalized autocorrelation function of Rouse modes,
$\langle X_p(t)X_p(0) \rangle / \langle X_p^2 \rangle$, versus time $t [\tau]$
for polymer melts of size $N=500$ and $N=2000$.
$N/p=5,10,25,50,125,250,\;{\rm and}\;500$ from left to right.
(b) $\ln[\langle X_p(t)X_p(0) \rangle / \langle X_p^2 \rangle]$
versus $tp^2/N^2[\tau]$ for $N=500$, and for several chosen values of $p$, as indicated.
(c) Same data for $N=500$ as shown in (a), but including the curves which present
the best fit of our data using Eq.~(\ref{eq-kww}) for comparison. All data are for $k_\theta=1.5$.}
\label{fig-xpxp-k1p5}
\end{figure*}

\begin{figure*}[t]
\resizebox{1.60\columnwidth}{!}{\includegraphics{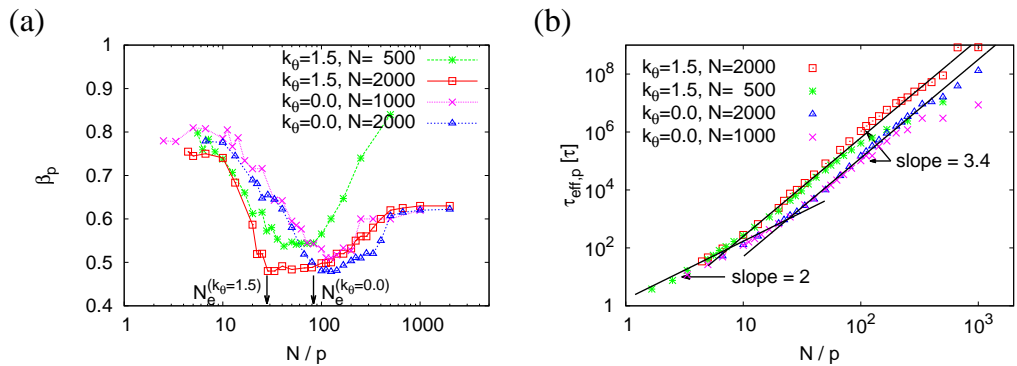}}
\caption{(a) Values of exponent $\beta_p$ from fitting the stretched
exponential function, Eq.~(\ref{eq-kww}), to the normalized
autocorrelation function of Rouse modes as shown in Fig.~\ref{fig-xpxp-k1p5} plotted
versus $N/p$. (b) Effective relaxation times $\tau_{{\rm eff},p}$ plotted
as a function of $N/p$.
In (a) $N/p \approx N_e$ are indicated by arrows for $k_\theta=0.0$ and $1.5$.
In (b) two scaling laws $\tau_{{\rm eff},p} \sim (N/p)^2$ and $\tau_{{\rm eff},p} \sim (N/p)^{3.4}$
predicted by Rouse model and reptation theory, respectively, are also shown
for comparison.}
\label{fig-taubeta}
\end{figure*}

\begin{figure*}[t!]
\resizebox{1.60\columnwidth}{!}{\includegraphics{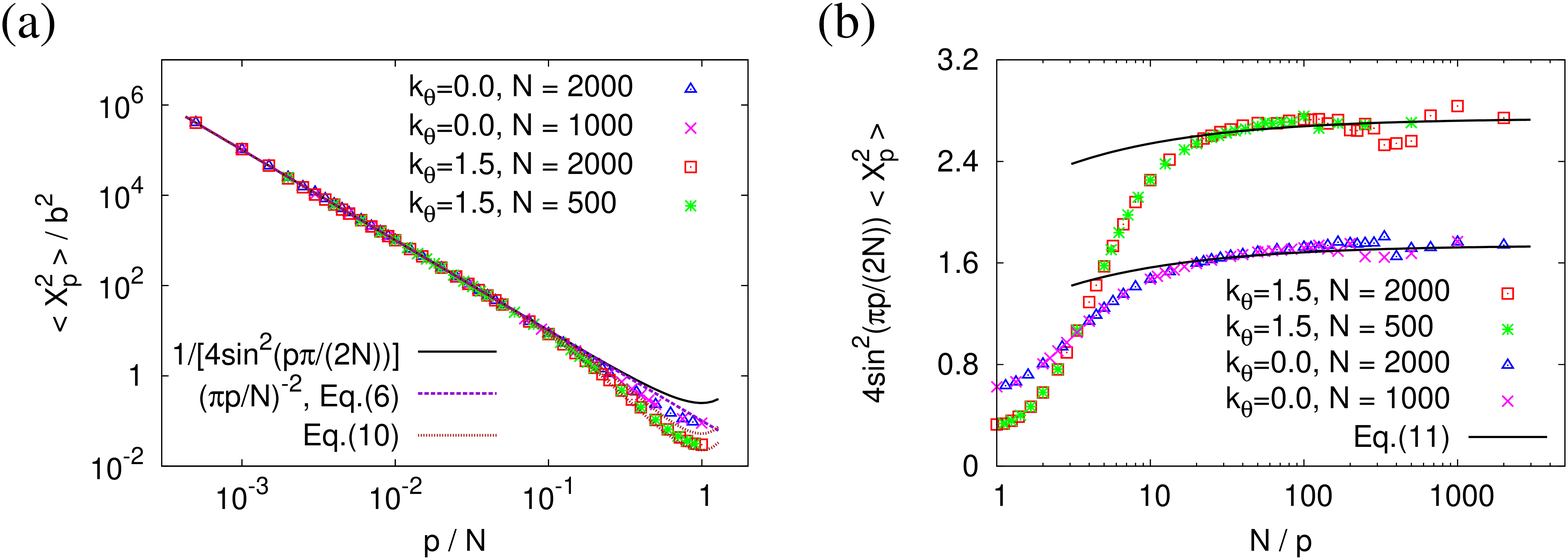}}
\caption{Rescaled amplitude of the autocorrelation function of the Rouse modes,
$\langle X_p^2 \rangle/b^2$ (a) and $4\sin^2(\pi p/(2N))\langle X_p^2 \rangle$ (b),
plotted versus $p/N$ and $N/p$ respectively.
Two chain sizes $N=500$, $2000$ are chosen for $k_\theta=1.5$,
and $N=1000$, $2000$ for $k_\theta=0.0$, as indicated.
In (a), theoretical predictions given in Eqs.~(\ref{eq-xpxp0}) and (\ref{eq-xpxp0-new})
are shown for comparison. $b^2=2.74 \sigma^2$ for $k_\theta=1.5$ and $b^2=1.74 \sigma^2$ for $k_\theta=0.0$.
In (b), Eq.~(\ref{eq-xpxp0-c}) with $C=0.23$ and $0.32$ for $k_\theta=1.5$ and $k_\theta=0$, respectively,
are also shown for comparison.}
\label{fig-xp0}
\end{figure*}

Figure~\ref{fig-xpxp-k1p5} shows the typical relaxation of
the time-dependent normalized autocorrelation function of modes $p$,
$\langle X_p(t)X_p(0) \rangle / \langle X_p^2 \rangle$,
according to the definition of Rouse modes $X_p(t)$ given in Eq.~(\ref{eq-xpdef}).
Data are for polymer melts of chain sizes $N=500$ and $2000$ with $k_\theta=1.5$.
The data sets shown in Fig.~\ref{fig-xpxp-k1p5}(a) from left to right correspond to 
$N/p=5,10,25,50,125,250,{\rm and}\;500$.
We see that for $N/p<N_e$ ($N_e \approx 28$ for $k_\theta=1.5$)
the relaxation of $\langle X_p(t)X_p(0) \rangle / \langle X_p^2 \rangle$ is independent 
of chain size $N$, namely, data for $N=500$ and $N=2000$ are on top of each other.
Similar results are also observed for polymer melts of chain sizes $N=1000$ and $2000$ with
$k_\theta=0.0$ where the entanglement length $N_e \approx 87$~\cite{Moreira2015} (not shown).
In the regime where $N/p>N_e$ the deviation between two data sets corresponding to
the same value of $N/p$ becomes more prominent as $N/p$ increases since the entanglement 
effect between chain segments becomes more important.
In Fig.~\ref{fig-xpxp-k1p5}(b), the plot of $\ln[\langle X_p(t)X_p(0) \rangle / \langle X_p^2 \rangle]$
versus $t/\tau_p$ with $\tau_p=(N/p)^2$ (see Eq.~(\ref{eq-Rouse}))
shows that the exponential decay is only valid for small values of $N/p$ at initial relaxation time $t$.
As $p$ becomes small and $t$ increases, one sees systematic deviations from the exponential decay due to the 
crossover from Rouse to reptation behavior.
In Fig.~\ref{fig-xpxp-k1p5}(c), the curves indicate the best fit of our data for 
$N=500$ to the theoretical prediction \{Eq.~(\ref{eq-kww})\} with
two fitting parameters $\beta_p$ and $\tau_{p}^*$.
Polymer chains containing $N/p$ monomers are relaxed completely
as $\langle X_p(t)X_p(0) \rangle / \langle X_p^2 \rangle \rightarrow 0$ for $t>>1$.
Therefore, one can estimate roughly the required relaxation time  
to relax very long chains in a melt through this curve fitting procedure.
Fitted values of the stretching exponent $\beta_p$ and
the estimates of the effective relaxation time $\tau_{{\rm eff},p}$ from $\beta_p$ and $\tau_p^*$
\{Eq.~(\ref{eq-taueff})\} are shown in Fig.~\ref{fig-taubeta}.
We see that $\beta_p$ reaches a minimum around $N/p \approx N_e$ for 
both cases $k_\theta=1.5$ and $0.0$ in Fig.~\ref{fig-taubeta}a. 
For chains having the same
intrinsic stiffness, $\beta_p$ is independent of size $N$ for
$N/p<N_e$ while $\beta_p$ decreases with increasing size $N$ for $N>N_e$.
Our results agree with the argument that the development of
a minimum in $\beta_p$ is due to kinetic constraints on the 
chains~\cite{Shaffer1995,Padding2002}, and the recent work~\cite{Kalathi2014}
using the same simulation model but different cut-off for the LJ potential.
At short length scales,
the local constraints on the motion of monomers is related to the chain connectivity
and the excluded volume interactions between monomers while at long
length scales, the entanglement effect sets in that the motion of entangled
chains is strongly hindered by topological constraints.
Results of the effective relaxation time $\tau_{{\rm eff},p}$ show 
the crossover behavior from the Rouse regime ($\tau_{{\rm eff},p} \sim (N/p)^2$)
to the reptation regime ($\tau_{{\rm eff},p} \sim (N/p)^{3.4}$) as $N/p$ increases.

In previous Monte Carlo simulations of polymer melts based on the bond 
fluctuation model~\cite{Kreer2001}, the authors showed that  
the Rouse model overestimates the correlation as $p/N>{\cal O}(10^{-1})$
due to the lack of considering the intrinsic stiffness of the chains.
Replacing random walk chains by freely rotating chains with a 
specific bond angle $\theta$,
the analytical expression of the amplitude $\langle X_p(0)X_p(0) \rangle$ 
is as follows,
\begin{widetext}
\begin{eqnarray}
    \langle \vec{X}_p(0) \vec{X}_p(0) \rangle 
 = b^2
\left\{ \left[4 \sin^2 \left(\frac{p\pi}{2N} \right) \right]^{-1}
- \left[\frac{1-\mid \langle \cos \theta \rangle \mid^2}
{4\mid \langle \cos \theta \rangle \mid} 
+4 \sin^2 \left(\frac{p\pi}{2N} \right) \right]^{-1}(1+{\cal O}(N^{-1}) \right\} \,.
\label{eq-xpxp0-new}
\end{eqnarray}
\end{widetext}

Figure~\ref{fig-xp0}a presents the rescaled amplitude of the autocorrelation function
of Rouse mode $p$, $\langle X_p(0) X_p(0) \rangle/b^2$, plotted versus $p/N$.
Here $b^2$ is determined by the best fit of our data to Eq.~(\ref{eq-xpxp0}) for small values of $p/N$.
$b^2=2.74 \sigma^2$ for $k_\theta=1.5$ and $b^2=1.74 \sigma^2$ for $k_\theta=0.0$.
Our fitted values of $b^2$ for both cases satisfy the relation $b^2=\ell_b^2 C_\infty$ 
predicted for freely rotating chains, where $C_\infty$ is Flory's characteristic ratio
($C_\infty=2.88$, $1.83$ for $k_\theta=1.5$, $0.0$, respectively), and
$\ell_b=0.964 \sigma$ is the mean bond length~\cite{Hsu2016,Moreira2015}.
Theoretical predictions given in Eqs.~(\ref{eq-xpxp0}) and (\ref{eq-xpxp0-new})
are also shown for comparison. The deviation from the Rouse prediction for
$p/N>{\cal O}(10^{-1})$ is indeed seen as shown in Ref.~\cite{Kreer2001}.
Taking the estimates of $\langle \cos \theta \rangle$
from our simulations, our data are described quite well by Eq.~(\ref{eq-xpxp0-new}) 
for $p/N>{\cal O}(10^{-1})$.
For small $p/N$ (large $N/p$), one should expect that $4\sin^2(p\pi/(2N))\langle X_p^2 \rangle$
reaches a plateau value $b^2$. However, since at short length scale the local intramolecular correlations in
the chains (the correlation hole effect) are important, a correction term~\cite{Wittmer2007,Semenov2010,Kalathi2014} 
${\cal O}((N/p)^{-1/2})$ is needed to be considered as follows,
\begin{equation}
      4\sin^2(p\pi/(2N))\langle X_p^2 \rangle = b^2[(1-C(N/p)^{-1/2}]
\label{eq-xpxp0-c}
\end{equation}
where $C$ is a fitting parameter.
The prediction is also verified as shown in Fig.~\ref{fig-xp0}b.
Since the entanglement effect already sets in at $N/p \approx 28$ for $k_\theta=1.5$,
we see that for chains of size $N=2000$, the data for $k_\theta=1.5$ fluctuate more than that for $k_\theta=0.0$.

\begin{figure*}[t!]
\resizebox{1.60\columnwidth}{!}{\includegraphics{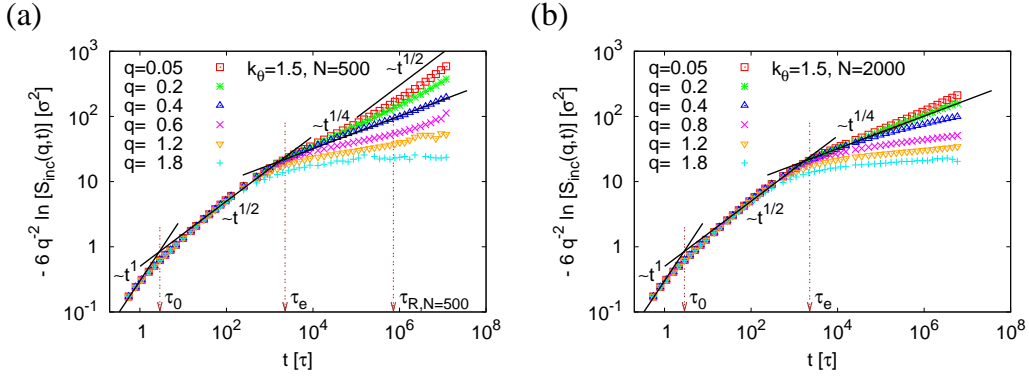}}
\caption{Log-log plot of $-6q^{-2}\ln S_{\rm inc}(q,t)$
versus $t$ for polymer melts of
sizes $N=500$ (a) and $N=2000$ (b).
Data are for polymer melts with $k_\theta=1.5$. The theoretical prediction
of the scaling behavior of the mean square displacement of monomers
is also shown for comparison since
$\ln S_{\rm inc}(q,t) \propto g_1(t)$. Here the characteristic relaxation
time $\tau_0 \approx 2.89 [\tau]$, the entanglement time
$\tau_e = \tau_0N_e^2 \approx 2266[\tau]$, and the Rouse time
$\tau_{R,N=500} = \tau_0N^2 \approx 7.2\times 10^5 [\tau]$ for $N=500$ are taken
from Ref.~\cite{Hsu2016}.}
\label{fig-sqinc}
\end{figure*}

\section{Dynamic structure factors}
 Dynamic behavior of polymer chains in a melt 
can also be described by the dynamic scattering from single chains.
The coherent and incoherent dynamic structure factors 
$S_{\rm coh}(q,t)$ and $S_{\rm inc}(q,t)$ for single chains are defined by 
\begin{equation}
  S_{\rm coh}(q,t)= \frac{1}{N} \langle
\sum_{i=1}^N \sum_{j=1}^N \exp\{i\vec{q} \cdot [\vec{r}_i(t)-\vec{r}_j(0)]\} 
\rangle \,,
\label{eq-sqcoh}
\end{equation}
and
\begin{equation}
  S_{\rm inc}(q,t)= \frac{1}{N} \langle 
\sum_{i=1}^N \exp\{i \vec{q} \cdot [\vec{r}_i(t)-\vec{r}_i(0)]\}
\rangle \,.
\label{eq-sqinc}
\end{equation}
The average $\langle \cdots \rangle$ denotes an average over all chains,
many starting states $(t=0)$, as well as over orientations of 
the wave vector $\vec{q}$ having the same wave length.
Note that $S_{\rm coh}(q)$ is the $q$-space representation
of the Rouse modes.

In the Rouse model, the displacement between monomer positions 
is Gaussian distributed since the force has a Gaussian probability
distribution. Therefore, Eqs.~(\ref{eq-sqcoh}) and ~(\ref{eq-sqinc})
can be written as
\begin{equation}
  S_{\rm coh}(q,t)= \frac{1}{N} 
\sum_{i=1}^N \sum_{j=1}^N \exp\{-\frac{1}{6}q^2
\langle [\vec{r}_i(t)-\vec{r}_j(0)]^2 \rangle \}  \,,
\label{eq-sqcoh1}
\end{equation}
and
\begin{equation}
  S_{\rm inc}(q,t)= \frac{1}{N}  
\sum_{i=1}^N  \exp\{-\frac{1}{6}q^2 \langle 
[\vec{r}_i(t)-\vec{r}_i(0)]^2 \rangle \} \,.
\label{eq-sqinc1}
\end{equation}
where $\langle [\vec{r}_i(t)-\vec{r}_i(0)]^2 \rangle \sim g_1(t)$ is simply the mean
square displacement of monomers.

For short chains ($N<N_e$), the scaling predictions of $S_{\rm coh}(q,t)$ and 
$S_{\rm inc}(q,t)$ for $t<\tau_e$ ($N<N_e$) and $q>2 \pi/R_g (N)$ ($R_g(N)$ is the radius of 
gyration of chains containing $N$ monomers) predicted by
the Rouse model are as follows,
\begin{equation}
 \ln [S_{\rm coh}(q,t)/S_{\rm coh}(q,0)] = -q^2(Wt)^{1/2}/6 \,,
\label{eq-sqcoh-rouse}
\end{equation}
and 
\begin{equation}
 \ln S_{\rm inc}(q,t) = -q^2(Wt)^{1/2}/6 
\label{eq-sqinc-rouse}
\end{equation}
where the factor $W=\frac{12k_BTb^4}{\pi \zeta}$ and $R_g$ is the radius of gyration
of the chain of size $N$.
For $t \gg \tau_e$ one expects the standard diffusion behavior, i.e.
\begin{equation}
  \ln [S_{\rm coh}(q,t)/S_{\rm coh}(q,0)]  =q^2Dt
\end{equation}

\begin{figure*}[t!]
\resizebox{1.60\columnwidth}{!}{\includegraphics{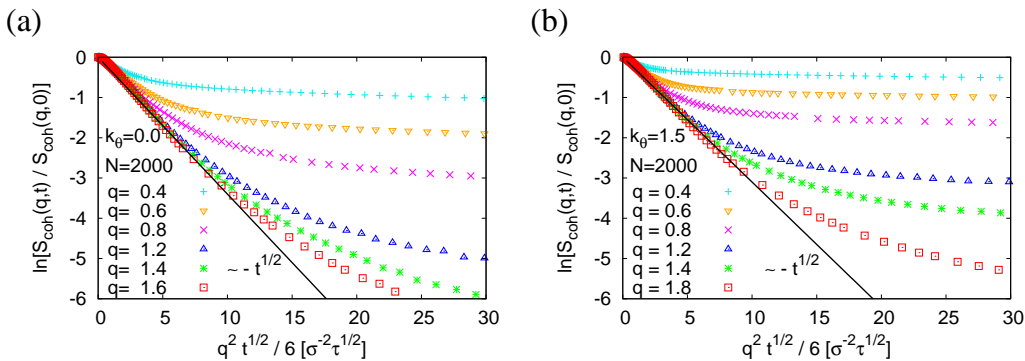}}
\caption{$\ln S_{\rm coh}(q,t)/S_{\rm coh}(q,0)$ plotted versus $q^2t^{1/2}/6 
[\sigma^{-2}\tau^{1/2}]$ for
polymer melts of size $N=2000$ and for $k_\theta=0.0$ (a) and $k_\theta=1.5$ (b).
The scaling law predicted by the Rouse model, $t^{1/2}$ is shown by a straight line.}
\label{fig-lsqcoh}
\end{figure*}

For long chains ($N>N_e$), the entanglement effect due to the topological
constraints between chains in a melt becomes important. According
to the reptation theory, local
reptation processes for short time and escape processes from the tube (creep motion)
for longer times and small values of $q$ should be considered.
Thus a pronounced plateau in $S_{\rm coh}(q,t)$ is predicted and
can loosely be interpreted as a Debye-Waller factor for $\tau_e \ll t \ll \tau_d$,
\begin{equation}
    \frac{S_{\rm coh}(q,t)}{S_{\rm coh}(q,0)} = 1-q^2d^2/36 \,.
\label{eq-sqcoh-rep}
\end{equation}
Note that here the tube diameter is defined
by $d=R_e(N_e)$ where $R_e(N_e)$ is the end-to-end
distance of chains of size $N_e$~\cite{Doi1986,Puetz2000}. In our previous work~\cite{Hsu2016},
the definition of tube diameter $d_T$ is different by a factor of $\sqrt{3}$, i.e. our simulation estimate of
tube diameter $d_T=(2 \langle R_g^2(N_e) \rangle)^{1/2}=(\langle R^2_e(N_e) \rangle/3)^{1/2}=d/\sqrt{3}$.
In the deep-reptation regime,
an analytic expression of the coherent dynamic structure used often
in the neutron-spin-echo (NSE) measurements for the determination of
the tube diameter is given by~\cite{Doi1986,Kremer1990,Puetz2000,deGennes1981,kremer1984}
\begin{eqnarray}
  \frac{S_{\rm coh}(q,t)}{S_{\rm coh}(q,0)} =
\left\{\left[1-\exp\left(-\frac{q^2 d^2}{36}\right) \right] f(q^2(Wt)^{1/2}) + \right . \nonumber \\
\left .
+\exp \left[-\frac{q^2 d^2}{36}\right] \right\} \frac{8}{\pi^2} 
\sum_{n =1,odd}^\infty \frac{\exp[-t n^2/\tau_d]}{p^2} 
\label{eq-dtube}
\end{eqnarray}
with $f(u)=\exp(u^2/36){\rm erfc}(u/6)$.

Figure~\ref{fig-sqinc} shows the results of the incoherent dynamic structure factor 
$S_{\rm inc}(q,t)$ according to Eqs.~(\ref{eq-sqinc}) and (\ref{eq-sqinc-rouse}) 
for polymer melts of sizes $N=500$ and $2000$ with $k_\theta=1.5$. 
The characteristic time scales, $\tau_0$, $\tau_e$,
and  $\tau_{R,N=500}$ taken from Ref.~\cite{Hsu2016} are indicated 
by arrows.
Since $\ln S_{\rm inc}(q,t) \sim g_1(t)$, the scaling laws
of $g_1(t)$ showing the crossover behavior from the Rouse regime to the reptation 
regime as $t$ increases are also shown for comparison. 
In the Rouse regime where $q>2\pi/d_T \approx 1.25\sigma^{-1}$,
we see that
$\ln S_{\rm inc}(q,t) \sim t$ for $t<\tau_0$ while $\ln S_{\rm inc}(q,t) \sim t^{1/2}$
for $\tau_0<t<\tau_e$.
In the reptation regime one should expect that $\ln S_{\rm inc}(q,t) \sim t^{1/4}$ 
for $\tau_e<t<\tau_R$ and $2 \pi/R_g(N) < q <2 \pi/d_T$. 
We see that indeed $S_{\rm inc} \sim t^{1/4}$ for $0.4\sigma^{-1}<q<1.25\sigma^{-1}$ in Fig.~\ref{fig-sqinc}a
and for $0.2\sigma^{-1}<q<1.25\sigma^{-1}$ in Fig.~\ref{fig-sqinc}b. 
Here for $k_\theta=1.5$, $R_g \approx \sqrt{0.4839N}\sigma$~\cite{Hsu2016}. 
In Fig.~\ref{fig-sqinc}a, we have also observed that $\ln S_{\rm inc}(q) \sim t^{1/2}$ for 
$t>\tau_R$ and $q<0.4\sigma^{-1}$. Therefore, our results are in perfect agreement with the 
theoretical predictions.

\begin{figure*}[t!]
\resizebox{1.60\columnwidth}{!}{\includegraphics{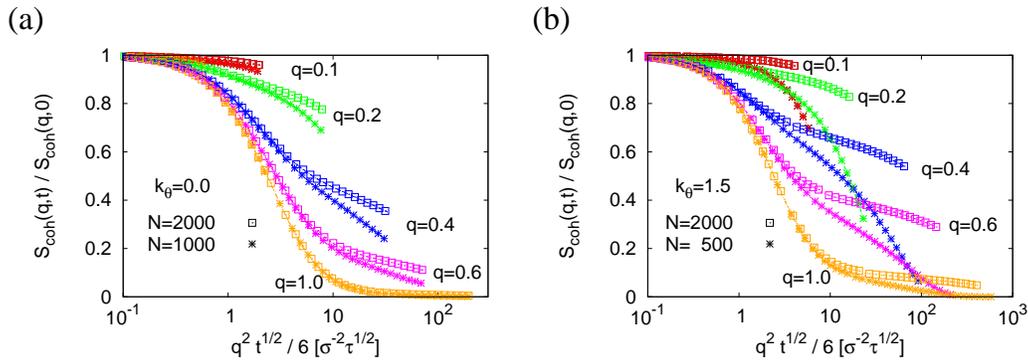}}
\caption{Semi-log plot of normalized coherent dynamic structure factor,
$S_{\rm coh}(q,t)/S_{\rm coh}(q,0)$, versus $q^2t^{1/2}/6$ for $k_\theta=0.0$ (a)
and $k_\theta=1.5$ (b). Five values of $q$ and two chain sizes $N$ are chosen, as indicated.}
\label{fig-sqcoh}
\end{figure*}

\begin{figure*}[t!]
\resizebox{1.60\columnwidth}{!}{\includegraphics{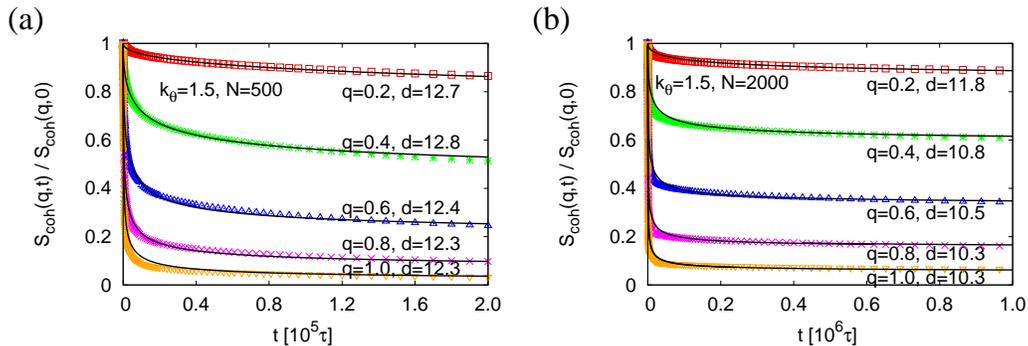}}
\caption{Normalized coherent dynamic structure factor,
$S_{\rm coh}(q,t)/S_{\rm coh}(q,0)$, plotted versus $t$ for polymer melts
of sizes $N=500$ (a) and $2000$ (b) with $k_\theta=1.5$.
Five values of $q$ are chosen, as indicated. The curves show the best fit 
of our data by adjusting the fitting parameter $d$ using Eq.~(\ref{eq-dtube}).}
\label{fig-fsqcoh}
\end{figure*}

For checking the scaling behavior of the normalized coherent dynamic structure
factor predicted by the Rouse model, we plot $\ln[S_{\rm coh}(q,t)/S_{\rm coh}(q,0)]$
versus $q^2 t^{1/2}/6$ for polymer melts of size $N=2000$ with $k_\theta=0.0$ and $1.5$
in Fig.~\ref{fig-lsqcoh}. We see that in the intermediate time regime $\tau_0<t<\tau_e$,
our data are also in perfect agreement with the scaling law $t^{1/2}$ for 
$q>2\pi/d_T \propto N_e^{-1/2}\sigma^{-1}$. 
For $\tau_d\gg t \gg \tau_e$, one would expect that $S_{\rm coh}(q,t)/S_{\rm coh}(q,0)$ 
reaches a plateau predicted by the reptation theory \{Eq.~(\ref{eq-sqcoh-rep})\}.
Therefore, in Fig.~\ref{fig-sqcoh}, we show the similar data as shown in Fig.~\ref{fig-lsqcoh} but
plot $S_{\rm coh}(q,t)/S_{\rm coh}(q,0)$ versus $q^2t^{1/2}/6$.
We see that first data tend to collapse onto a single master curve for $\tau_0<t<\tau_e$
as $q$ increases and $S_{\rm coh}(q,t)/S_{\rm coh}(q,0)$ is independent of chain size $N$.
As $t> \tau_e$, $S_{\rm coh}(q,t)/S_{\rm coh}(q,0)$ for two different chain sizes $N$ 
start to deviate from each other.
For polymer chains of size $N=2000$ in both cases ($k_\theta=0.0$ and $k_\theta=1.5$),
$S_{\rm coh}(q,t)/S_{\rm coh}(q,0)$ slows down as $t \gg \tau_e$.
It gives the first evidence from simulations that 
a pronounced plateau in $S_{\rm coh}(q,t)$ shall occur as chain size $N$ increases.
Finally, we determine the tube diameter $d_T=d/\sqrt{3}$ for $k_\theta=1.5$ 
by fitting our simulation data
of $S_{\rm coh}(q,t)/S_{\rm coh}(q,0)$ to Eq.~(\ref{eq-dtube}) (see Fig.~\ref{fig-fsqcoh}).
Our results show that $d_T\approx 7.10 \sigma$ for $N=500$, and
$d_T \approx 5.95 \sigma$ for $N=2000$ which are compatible to our previous
estimate of $d_T \approx 5.02 \sigma$~\cite{Hsu2016}.

\section{Conclusion}
  By extensive molecular dynamics simulations and accompanying theoretical
predictions in the literature, we have investigated the dynamic properties of polymer 
melts in equilibrium by analyzing the chain Rouse modes, and the dynamic
coherent and incoherent structure factors for chains of two different 
sizes and stiffnesses. 
The relaxation of time-dependent autocorrelation functions of Rouse modes $p$ is  
independent of chain size $N$ for $N/p<N_e$ (Fig.~\ref{fig-xpxp-k1p5}a) where the entanglement length
$N_e$ is obtained through the primitive path analysis (PPA)~\cite{Hsu2016,Moreira2015,Everaers2004}.
Estimates of the stretching exponent $\beta_p$ also show that the minimum 
of $\beta_p$ occurs in the vicinity of $N/p \approx N_e$ (Fig.~\ref{fig-taubeta}a). 
The crossover behavior of effective relaxation times $\tau_{\rm eff,p}$ from
the Rouse regime to the reptation chain as $N/p$ increases is verified as well.
Since all these estimated quantities for $N/p<N_e$ behave differently from that for $N/p>N_e$,
we see that $N_e$ can also be determined roughly via the Rouse mode analysis of large polymer
melt systems of two different chain sizes, and the value is
consistent with that obtained through PPA.
Our results are also in perfect agreement with the extended theoretical 
predictions considering the excluded interaction, topological
constraint, the intramolecular interactions, and chain stiffness (Fig.~\ref{fig-xp0}). 

  The scaling behavior of coherent and incoherent dynamic structure factors 
strongly depends on the time $t$ and wave length $q$.
Therefore, it is a delicate matter to analyze the dynamic structure factors.
However, we have provided evidence that the scaling behavior of $S_{\rm inc}(q,t)$
(Fig.~\ref{fig-sqinc})
is compatible with the mean square displacement of monomers $g_1(t)$, and 
the crossover points characterized by the characteristic time scales $\tau_0$, $\tau_e$,
and $\tau_R$ and the corresponding wave length scales (the inverse of length scales)
are consistent with each other. 
The slowing down of $S_{\rm coh}(q,t)$ (Fig.~\ref{fig-sqcoh}) gives the first evidence that 
$S_{\rm coh}(q,t)$ exhibits a plateau for $\tau_e \ll t \ll \tau_d$ as the chain size $N$ increases.
The tube diameter $d_T=d/\sqrt{3}$ extracted from $S_{\rm coh}$ (Fig.~\ref{fig-fsqcoh}) is also
in perfect agreement with $d_T$ obtained from $g_1(t)$ in Ref.~\cite{Hsu2016}.

   We hope that the present work showing the detailed analysis of the dynamic properties 
of highly entangled chains covering the scaling regimes from Rouse to reptation will help
for the further understanding of the dynamic behavior of the deformed polymer melts, 
polydisperse polymer melts, and the related experiments. \\

\underline{Acknowledgments:}
This work was supported by European Research Council under the European
Union's Seventh Framework Programme (FP7/2007-2013)/ERC Grant Agreement
No.~340906-MOLPROCOMP. We are grateful to G. S. Grest for stimulating discussions,
and A. C. Fogarty for a critical reading of the manuscript.
We are also grateful to the NIC J\"ulich for a generous
grant of computing time at the J\"ulich Supercomputing
Centre (JSC), and the Rechenzentrum Garching (RZG), the supercomputer center
of the Max Planck Society, for the use of their computers.
This work is dedicated to Wolfhard Janke on the occasion of his 60th birthday.
H.-P. Hsu is thankful to Wolfhard for the invitations to several scientific events and 
for the fruitful discussions during her annual visits to Leipzig.
%

\end{document}